\documentclass[seceq]{ptptex}

\usepackage{graphicx}
\def\ion#1#2{{\rm #1}\,{\small\sc{#2}}\relax}

\begin{document}
\markboth{Karen M. Leighly}{The SEDs of NLS1s}

\title{The Spectral Energy Distributions of Narrow-line Seyfert 1 Galaxies}

%\subtitle{Subtitle}    %use this when you want a subtitle

\author{%       %Use \scshape  for the family name
Karen M. \textsc{Leighly}}

\inst{%         %Affiliation, neglected when [addenda] or [errata]
Dept.\ of Physics and Astronomy, The University of Oklahoma, 440 W.\
Brooks St., Norman, OK 73019, USA
}

%\publishedin{%         %Write this ONLY in cases of addenda and errata
%Prog.~Theor.~Phys.\ \textbf{XX} (19YY), page.}

%\recdate{Mmmmm DD, YYYY}%            %Editorial Office will fill in this.

\abst{Narrow-line Seyfert 1 galaxies are identified by their uniform
  optical spectral properties.  Studies of samples of NLS1s reveal,
  however, a range of X-ray spectral and variability behavior, and UV
  spectral behavior. We describe the range of behavior observed,
  illustrating with a comparison of the {\it XMM-Newton} and {\it
  HST} data from the NLS1s 1H~0707$-$495 and Ton~S180, and discuss how
  the spectral energy distribution may causally link the UV and X-ray
  properties.}

\maketitle

\section{Introduction}

Narrow-line Seyfert 1 galaxies are identified by their optical
emission line properties.  They have narrow permitted optical lines
(FWHM of H$\beta<2000\rm\, km\,s^{-1}$), weak forbidden lines
([\ion{O}{III}]/H$\beta<3$; this distinguishes them from Seyfert 2
galaxies), and frequently they show strong \ion{Fe}{II}
emission.\cite{rf:1} In 1992, it was demonstrated by Boroson \&
Green\cite{rf:2} that the optical emission-line properties around
H$\beta$ are strongly correlated with one another. A principal
components analysis allowed the largest differences among optical
emission line properties to be gathered together in a construct
commonly known as ``Eigenvector 1''. The strongest differences hinge
on the strength of the \ion{Fe}{II} and [\ion{O}{III}] emission, and
the width and asymmetry of H$\beta$. These are just the properties
that define NLS1s.  During the 1990's, it was found that the X-ray
properties are also manifested in these correlations: NLS1s are
observed to have steeper soft X-ray spectra,\cite{rf:3} steeper hard
X-ray spectra,\cite{rf:4,rf:5} and higher amplitude X-ray
variability.\cite{rf:6} Properties of UV spectra also appear in
Eigenvector 1: NLS1s tend to have higher \ion{Si}{III}]/\ion{C}{III}]
ratios, stronger low-ionization lines, weaker \ion{C}{IV}, and
stronger \ion{N}{V}.\cite{rf:7}

These sets of strong correlations are remarkable, because they involve
dynamics and gas properties in emission regions separated by vast
distances. This pervasiveness leads us to believe that we are
observing the manifestation of a primary physical 
parameter. A favored explanation is that it is the accretion rate
relative to the black hole mass onto the active nucleus.  This is
easily understood from the X-ray variability properties as follows.
NLS1s have systematically higher fractional amplitude of variability
at a particular X-ray luminosity than do Seyfert 1 galaxies with broad
optical lines in {\it ASCA} observations (Fig.\ 1).\cite{rf:6} Since
the {\it ASCA} observations are all nearly the same length, a higher
fractional amplitude of variability implies a shorter variability time
scale.  A shorter time scale implies a smaller emission region, which
corresponds to a smaller black hole mass, assuming that the geometry,
etc., are uniform among Seyfert 1s.  Assuming a constant efficiency of
conversion of gravitational potential energy to radiation, the X-ray
luminosity corresponds to the absolute accretion rate.  Thus, for a
given luminosity or accretion rate, NLS1s have a smaller black hole
mass, and therefore have a higher accretion rate relative to the
Eddington value.

The high-accretion-rate scenario is attractive and simple.  However,
it appears to be incomplete.  In the process of analyzing the {\it
ASCA} X-ray spectra from NLS1s, I found that their properties did not
appear to be uniform.  Specifically, there is a correlation between
their fractional amplitude of variability and a parameter,
$\alpha_{xx}$, that measures the strength of the X-ray soft excess
(Fig.\ 1).\cite{rf:5} Objects with very prominent soft
excesses show high-amplitude variability, and objects with moderate
soft excesses show lower-amplitude variability.  Thus, a connection
between the shape of the X-ray spectrum and the variability among
NLS1s is observed.

\begin{figure}
\centerline{\includegraphics[width=12.2 cm]{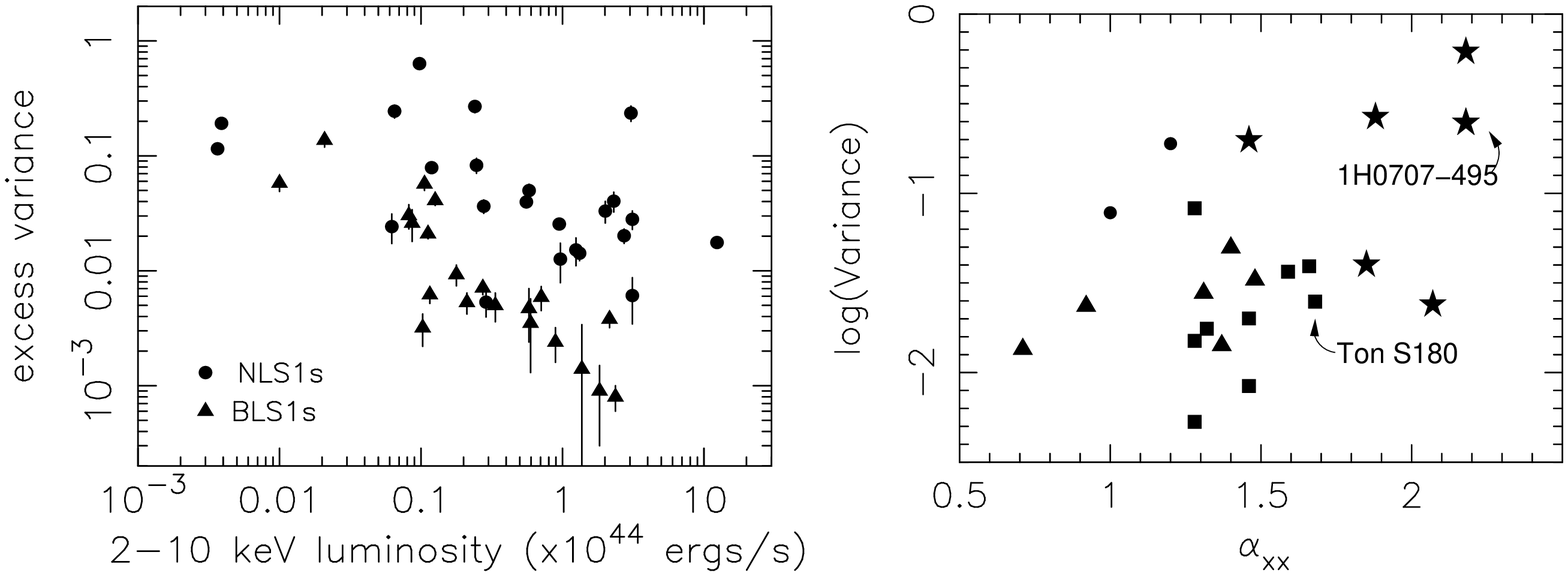}}
\caption{{\it Left:} Excess variance, defined as the measured variance
corrected for the measurement error and normalized by the mean, as a
function of the 2--10 keV luminosity of a sample of Seyfert 1 galaxies
observed with {\it ASCA}.\cite{rf:6}  {\it Right:} The excess
variance as a function of $\alpha_{xx}$, a parameter defined as the
point-to-point energy index between 0.7 and 4 keV of the best-fitting
continuum spectral model, for a sample of NLS1s.\cite{rf:5}}
\label{fig:1}
\end{figure}

The {\it ASCA} data were invaluable for understanding the properties
of NLS1s; however, because of the limited signal-to-noise ratio, we
could not investigate the nature of the soft excess or the variability
in detail. The large effective area of {\it  XMM-Newton} vastly
improves the situation.  We can gain some insight into the correlation
shown in Fig.\ 1 by looking at archival {\it XMM-Newton} data from two
NLS1s, 1H~0707$-$495 and Ton~S180, with different locations in the
correlation plot.

\section{X-ray Properties of 1H~0707$-$495 and Ton~S180}

1H~0707$-$495 and Ton S180 were observed using {\it XMM-Newton} for
$\sim 40$ and $\sim 30$ ks, respectively. We obtained the data from
the archive; both of these data sets have been
published.\cite{rf:8,rf:9}

Let's look first at the spectral properties.  The most notable feature
is that the character of the soft excess in these two objects is
completely different.  In 1H~0707$-$495, the soft excess is very
prominent and can be modelled very well by a single-temperature black
body (Fig.\ 2).  In contrast, the soft excess in Ton~S180 is shallower
and it requires two blackbodies to model it successfully; the spectrum
was modelled successfully using two Comptonized plasmas in
Ref.~\citen{rf:8}. These differences suggest that the process
responsible for the soft excess is different.

\begin{figure}
\centerline{\includegraphics[width=12.2 cm]{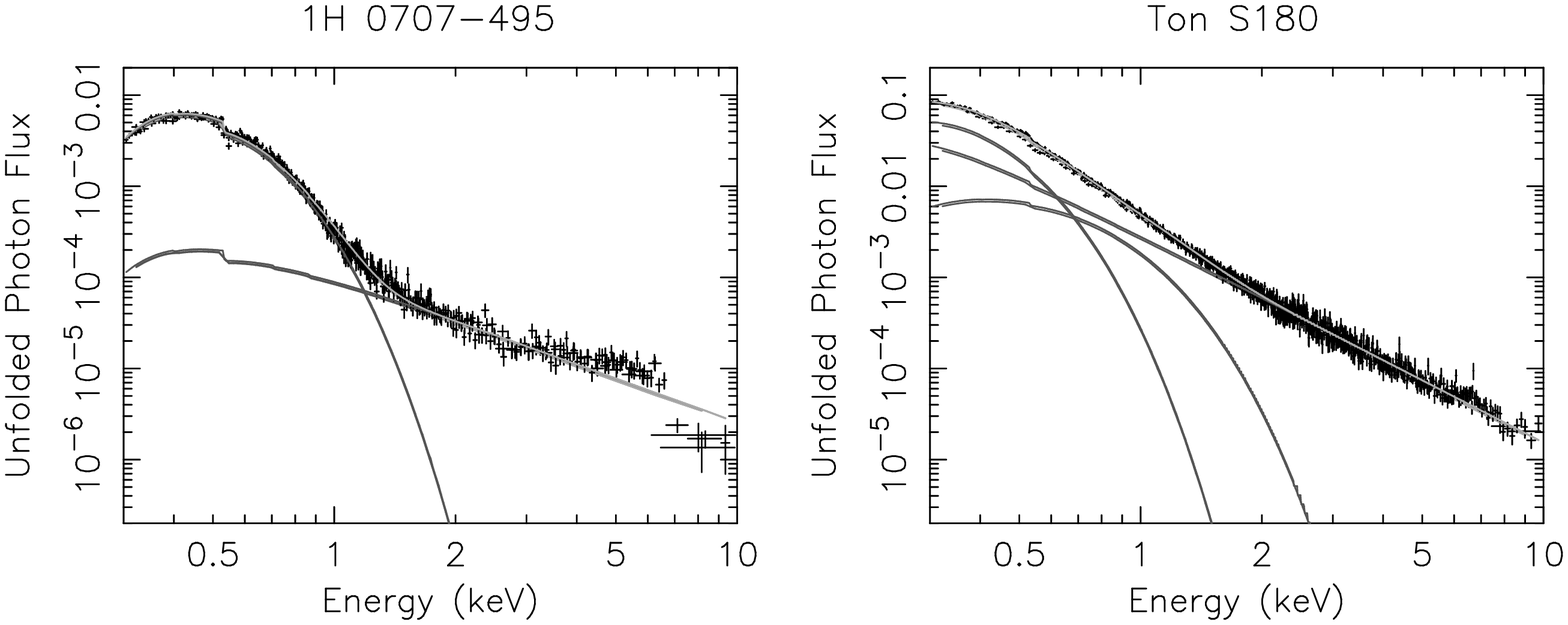}}
\caption{Continuum models for {\it XMM-Newton} spectra of two NLS1s,
  corrected for the detector response.  {\it Left:} The soft excess
  from 1H~0707$-$495 is very prominent and can be modeled as a single
  temperature blackbody. {\it Right:} The soft excess from Ton~S180
  is more subtle.  It is parameterized here as two blackbodies; the
  entire spectrum can also be modeled using two Comptonized
  plasmas.\cite{rf:8}} 
\label{fig:2}
\end{figure}

Turning next to the variability, we show in Fig.\ 3 the full-band
light curves.  We renormalized by dividing by the average count rate to
 emphasize the differences.  We can clearly see that
amplitude of variability is much higher in 1H~0707$-$495 than in
Ton~S180, as expected from the {\it ASCA} behavior.

\begin{figure}
\centerline{\includegraphics[width=12.2 cm]{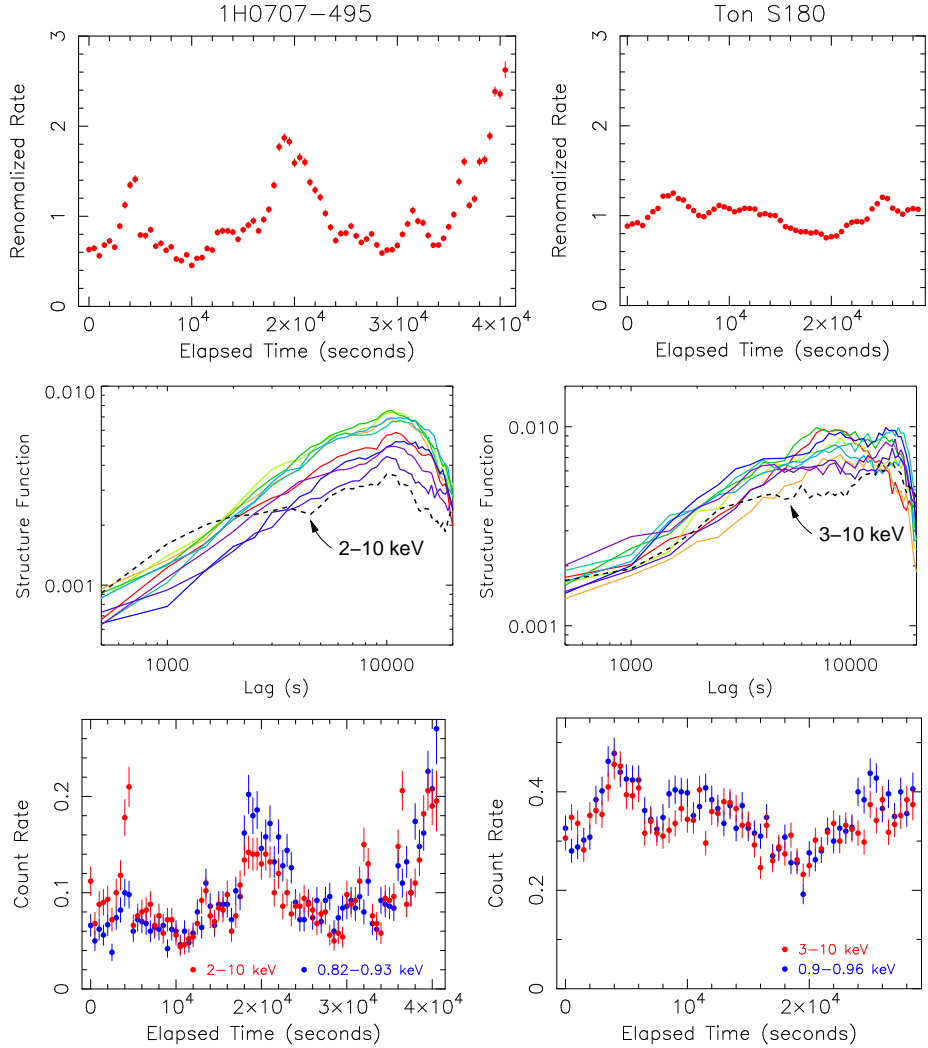}}
\caption{X-ray variability in the {\it XMM-Newton} observations of
  1H~0707$-$495 (left) and Ton~S180 (right).  {\it Top:} Full-band
  light curves renormalized so that they have a mean equal to 1.  The
  difference in variability amplitude is clearly seen.  {\it Middle:}
 The structure function of energy-sliced light curves; red is softest,
  black is hardest.  The difference in slope in the 2--10~keV
  band in 1H~0707$-$495 reveals more power on short time scales.  {\it
  Bottom:} Examination of the energy-sliced light curves shows that
  there are short time-scale high-amplitude flares in
  1H~0707$-$495.} 
\label{fig:3}
\end{figure}

It is often illuminating to examine at the variance as a function of
energy.  In these two observations, however, the variance is constant,
suggesting that there is little spectral variability.\cite{rf:8,rf:9}
The variance is the integral over the power spectrum from
$2\pi/T_{obs}$ to $2\pi/T_{bin}$, where $T_{obs}$ and $T_{bin}$ are the length
of the observation and the bin size, respectively. Performing the
integral averages over all time scales so that information about
variability on different time scales  is lost.  

Computing the variance
was about all one could do with the {\it ASCA} data, because of the low
signal-to-noise ratio. We can do more with the  {\it   XMM-Newton} data.  
Fig.\ 3 shows the structure function (SF) of energy-sliced light
curves.  The behavior of most of the SFs run parallel to one another,
indicating no spectral variability. The exception is the highest
energy band.  In 1H~0707$-$495, the 2--10 keV SF is flatter toward
short time scales than the SFs in the other bands, indicating more
power on shortest time scales. We can understand this result by
examining the energy-sliced light curves, shown in Fig.\ 3.  In
Ton~S180, the hard and soft light curves are nearly identical.  But in
1H~0707$-$495, several high-amplitude, short time-scale flares are seen
in the hard band.  It is worth noting that in the 2002 {\it Chandra}
observation of 1H~0707$-$495, similar behavior was observed, except
that the hard X-rays were much more variable on all time scales than
the soft X-rays.\cite{rf:10} 

My interpretation of these results is as follows.  In 1H~0707$-$495,
there is a soft excess and power law, and the difference in
variability implies that they come from two physically distinct,
although coupled, emission regions.  In Ton~S180, although the spectrum
is parameterized in terms of a soft excess and a power law, the
correlated variability suggests that these components are not
distinct; perhaps Comptonization is occurring in gas with clumps
having a range of optical depths and temperatures. 

\section{The UV Properties of NLS1s}

Turning now to the optical spectra, shown in Fig.\ 4, we see the
narrow H$\beta$, strong \ion{Fe}{II}, and weak [\ion{O}{III}]
characteristic of NLS1s.  There is little difference between
1H~0707$-$495 and Ton~S180, as expected.  The interesting differences
appear in the UV spectra.  In both cases, the broadband continuum is
similar.  In the far-UV spectrum we find that 1H~0707$-$495 shows
stronger \ion{N}{V}, and stronger \ion{Si}{II}. 
The most interesting differences are in the \ion{C}{IV} profile.  In
1H~0707$-$495, the line is strongly blueshifted, with the
largest part of the line blueward of the rest wavelength.  In
Ton~S180, while there is a blueshifted tail, the largest part of the
line is roughly centered at the rest wavelength.

\begin{figure}
\centerline{\includegraphics[width=12.2 cm]{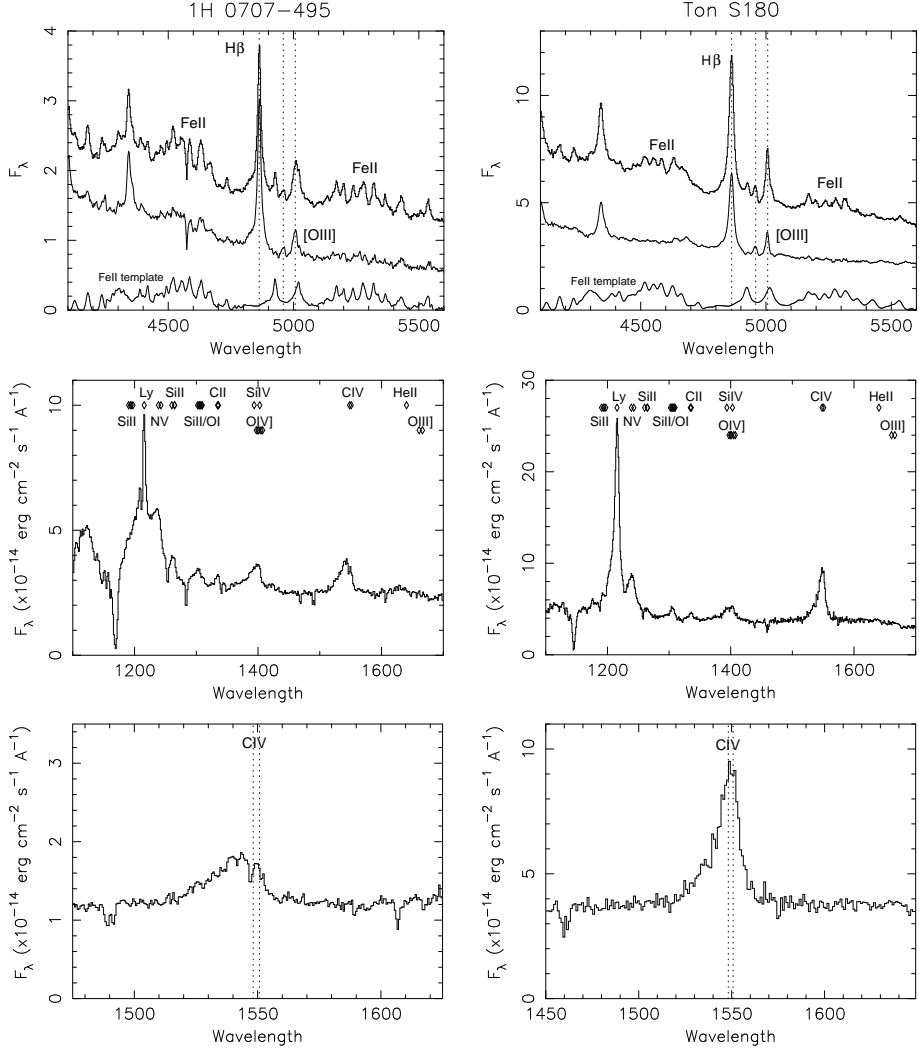}}
\caption{Optical and UV emission properties of 1H~0707$-$495 (left)
  and Ton~S180   (right).  {\it Top:} Both objects are classified as
  NLS1s; insignificant differences in the optical spectra are seen.
  {\it   Middle:} 1H~0707$-$495 is
  characterized   by relatively   stronger \ion{N}{V} and low-ionization
  lines such as   \ion{Si}{II}   than is Ton~S180.  Note that
  none of the   absorption lines are   intrinsic. {\it Bottom:} The
  \ion{C}{IV} line in 1H~0707$-$495 has lower equivalent width and is
  significantly blueshifted compared with the \ion{C}{IV} line in Ton~S180.} 
\label{fig:4}
\end{figure}

Considering a larger sample of 16 NLS1s, we see that the \ion{C}{IV}
differences are part of a trend (Fig.\ 5).\cite{rf:11} The \ion{C}{IV}
equivalent width is strongly anticorrelated with a parameter that
describes the asymmetry of the line which is defined as the fraction of the
emission blueward of the rest wavelength.  We suggest that \ion{C}{IV}
may be comprised of two components, a blueshifted portion that is
emitted in a wind, and a narrow part that is emitted in low-velocity
gas perhaps associated with the accretion disk or base of the wind.
In some NLS1s, such as 1H~0707$-$495, the \ion{C}{IV} line is
dominated by wind emission.

\begin{figure}
\centerline{\includegraphics[width=12.2 cm]{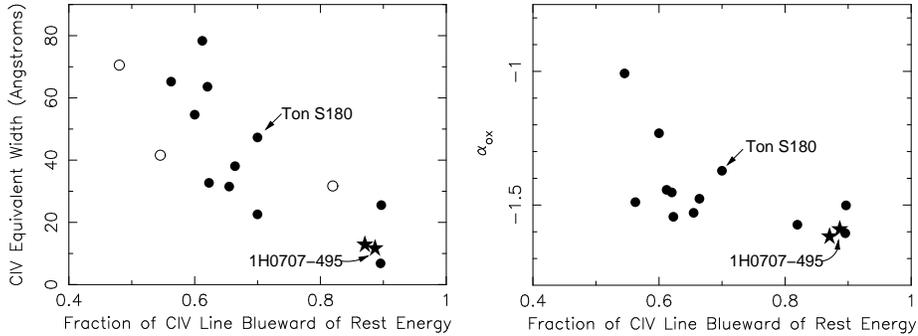}}
\caption{Broadband and emission line properties of a sample of
  NLS1s.\cite{rf:11} {\it Left:} The \ion{C}{IV} equivalent width as a
  function of the fraction of the line blueward of the rest energy.
  Objects with large values of the asymmetry parameter have very
  blueshifted lines; objects with values near 0.5 have very symmetric
  lines.   {\it Right:} $\alpha_{ox}$, the point-to-point slope
  between 2500\AA\/ and 2~keV, as a function of the asymmetry
  parameter.  Objects with strongly blueshifted lines are relatively
  X-ray weak.} 
\label{fig:5}
\end{figure}

What determines the presence of a wind?  There are theoretical
arguments and observational evidence that suggests that it is the
spectral energy distribution.  Fig.\ 5 shows
$\alpha_{ox}$\footnote{$\alpha_{ox}$ is defined as the point-to-point
energy index between 2500\AA\/ and 2 keV.} as a function of the
asymmetry parameter.\cite{rf:11} This shows that objects with
blueshifted lines are also X-ray weak. This makes perfect sense, if
the acceleration mechanism for the wind is radiative-line driving,
because strong UV emission is needed to drive the wind, while the
X-rays should be weak in order to prevent the wind from being
over-ionized.\cite{rf:12}  Fig.\ 6, showing the spectral energy
distributions of 1H~0707$-$495 and Ton~S180, reveals that the SED is
softer in 1H~0707$-$495.

\begin{figure}
\centerline{\includegraphics[width=12.2 cm]{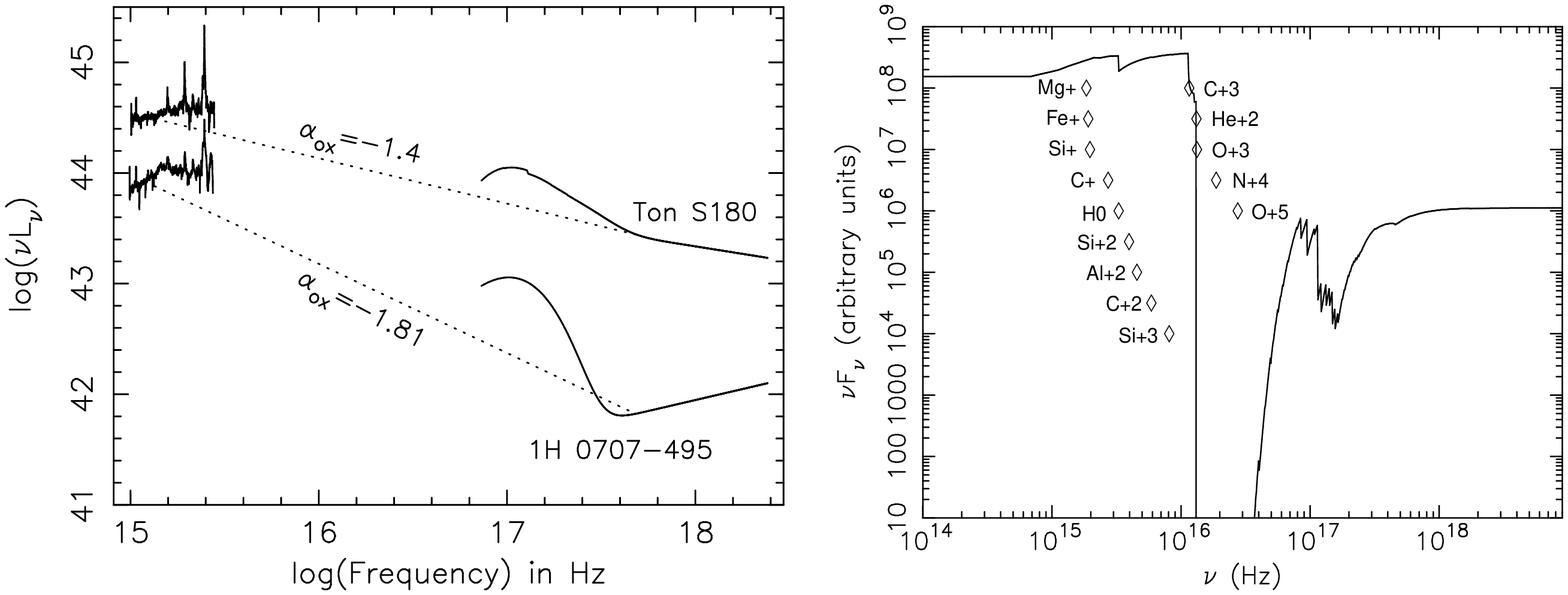}}
\caption{{\it Left:} The spectral energy distributions of
  1H~0707$-$495 and Ton~S180.  The {\it HST} spectra shown were scaled
  to the {\it XMM-Newton} OM fluxes. {\it Right:} The continuum after
  being transmitted through the wind lacks photons in the helium
  continuum.\cite{rf:14}} 
\label{fig:6}
\end{figure}

Let us next look at the intermediate- and low-ionization lines. One of
the peculiar properties of the UV spectra of NLS1s is the fact that
they contain strong very high ionization lines, such as \ion{N}{V},
and very strong low-ionization lines, such as \ion{Si}{II} and
\ion{Fe}{II}.  Also, the intermediate-ionization lines emitted tend to
be those with lower ionization potentials (i.e., \ion{Si}{III}] and
\ion{Al}{III} rather than \ion{C}{III}]).\cite{rf:11} This may have a
natural explanation in terms of the wind scenario described above. If
the continuum is transmitted through or is ``filtered''\footnote{We
differentiate between a ``shielded'' continuum, which is assumed to
have been transmitted through highly ionized gas,\cite{rf:13} and a
``filtered'' continuum, which is assumed to have been transmitted
through the wind while ionizing and exciting it before illuminating
the disk or low-velocity base of the wind and producing the observed
intermediate- and low-ionization 
lines.} by the wind before illuminating the
relatively dense, low-velocity gas that emits the intermediate- and
low-ionization lines, then it will lack photons beyond the helium edge
(Fig.\ 6) and will be unable to produce high-ionization line emission.
The intense continuum still excites the gas, causing it to emit strong
low-ionization lines, and lower I.P.\ intermediate-ionization
lines. This scenario and supporting photoionization modeling is
discussed in much more detail in Ref.~\citen{rf:14}.

We acknowledge that the anticorrelation between $\alpha_{ox}$ and the
emission line asymmetry is not very strong.  Most of the X-ray and UV
observations used to determine $\alpha_{ox}$ were not simultaneous, a
fact that would tend to reduce the correlation.  However, we find the
same results in several objects studied in detail.  Coordinated {\it
FUSE} and {\it ASCA} observations of RE~1034+39 revealed flat
$\alpha_{ox}$ of $-1.2$; this object is well known for its very hard
spectral energy distribution.\cite{rf:14} A hard spectral energy
distribution should yield narrow high-ionization lines centered at the
rest wavelength; no wind is present because the strong X-ray emission
overionizes the gas. Indeed, we find that all of the emission lines,
including the very high-ionization line \ion{O}{VI}, can be modeled
with nearly the same narrow profile.\cite{rf:16} In contrast,
coordinated {\it HST} and {\it Chandra} observations of the luminous
NLS1 PHL~1811 reveal a very soft, intrinsically X-ray weak
continuum.\cite{rf:17} A soft spectral energy distribution should
produce blueshifted high-ionization emission lines, and indeed, that
is what we find.

\section{Discussion}

In this talk, I presented a discussion of the X-ray spectral and
variability properties exhibited by NLS1s, drawing from studies of
{\it ASCA} spectra,\cite{rf:5,rf:6} and UV properties, drawing from
studies of {\it HST} spectra.\cite{rf:11,rf:14}  I showed that although
NLS1s are identified by fairly uniform optical spectral properties,
they exhibit a {\it range} of X-ray and UV spectral behaviors. 

I also presented an explicit comparison of X-ray and UV properties of
two NLS1s, 1H~0707$-$495 and Ton~S180.  These two objects show
distinctly different X-ray spectral and variability behaviors.  This
suggests that something about their central engines is different;
perhaps there are differences in geometry resulting from different
accretion rate.  There are also differences in their UV
properties; here I discussed the differences in the \ion{C}{IV}
profile, but note that there are other patterns in the range of
behaviors exhibited by NLS1s that are discussed in detail in
Ref.~\citen{rf:11}.  

The spectral analysis and photoionization modeling presented in
Ref.~\citen{rf:14} indicates that the shape of the spectral energy 
distribution may be responsible for the range of UV spectral
behaviors.  Specifically, when the UV continuum is blue and strong,
and the X-ray continuum is weak, a resonance-line driven wind is produced
that contributes blueshifted high-ionization emission lines.  This
wind filters the continuum before it illuminates the intermediate- and
low-ionization line-emitting region, causing it to emit strong
low-ionization lines, and lower I.P.\ intermediate-ionization lines.
In contrast, when the X-rays are strong relative to the UV, the gas
that would form the wind is overionized, and the resonance-line
driving fails. In that case, the high-ionization lines are narrow and
centered at the rest wavelength.  

The spectral energy distribution is produced in the central engine.
X-ray properties are key for understanding the central engine, because
the X-rays are emitted very close to the black hole.  I propose that the
differences seen in the X-ray properties of 1H~0707$-$495 and Ton~S180
are manifested in their spectral energy distributions, and perhaps
also the geometry of the central engine, which then influences their
UV emission line properties. This suggests a chain of causality
linking the X-ray and UV behaviors perhaps arising from a range in
some intrinsic parameter among members of the class of NLS1s. 

\section*{Acknowledgements}
KML would like to thank Andrea Crews, Chiho Matsumoto, John Moore, and
Darrin Casebeer for contributions to the work presented here.

%\appendix
%\section{First Appendix} %Empty argument \section{} yields `Appendix'. 
%
%\section{Second Appendix}


\begin{thebibliography}{99}
%%%%%%%%%%%%%%%%%%%%%%%%%%%%%%%%%%%%%%%%%%%%%%%%%%%%%%%%%%%%%
% Some macros are available for the bibliography:
%  o for general use
%    \JL : general journals                 \andvol : Vol (Year) Page
%  o for individual journal 
%    \AJ   : Astrophys. J.           \NC         : Nuovo Cim.
%    \ANN  : Ann. of Phys.           \NPA, \NPB  : Nucl. Phys. [A,B]
%    \CMP  : Commun. Math. Phys.     \PLA, \PLB  : Phys. Lett. [A,B]
%    \IJMP : Int. J. Mod. Phys.      \PRA - \PRE : Phys. Rev. [A-E]     
%    \JHEP : J. High Energy Phys.    \PRL        : Phys. Rev. Lett.
%    \JMP  : J. Math. Phys.          \PRP        : Phys. Rep.
%    \JP   : J. of Phys.             \PTP        : Prog. Theor. Phys.     
%    \JPSJ : J. Phys. Soc. Jpn.      \PTPS       : Prog. Theor. Phys. Suppl.
% Usage:
%  \PRD{45,1990,345}          ==> Phys.~Rev.\ \textbf{D45} (1990), 345
%  \JL{Nature,418,2002,123}   ==> Nature \textbf{418} (2002), 123
%  \andvol{B123,1995,1020}    ==> \textbf{B123} (1995), 1020
%%%%%%%%%%%%%%%%%%%%%%%%%%%%%%%%%%%%%%%%%%%%%%%%%%%%%%%%%%%%%
  
\bibitem{rf:1}
   D.\ E.\ Osterbrock \& R.\ W.\ Pogge, \AJ{108,1985,187}
\bibitem{rf:2}
   T.\ A.\ Boroson \& R.\  F.\ Green, \JL{ApJS,80,1992,109}
\bibitem{rf:3}
   T.\ Boller, W.\ N.\ Brandt, \& H.\ Fink, \JL{A\&A,279,1993,53}
\bibitem{rf:4}
   W.\ N.\ Brandt, S.\ Mathur, \& M.\ Elvis, \JL{MNRAS,285,1997,L25}
\bibitem{rf:5}
   K.\ M.\ Leighly, \JL{ApJS,125,1999,317}
\bibitem{rf:6}
   K.\ M.\ Leighly, \JL{ApJS,125,1999,297}
\bibitem{rf:7}
   B.\ J.\ Wills, A.\ Laor, M.\ S.\ Brotherton, D.\ Wills, B.\ J.\
   Wilkes, G.\ J.\ Ferland, \& Z.\ Shang, \JL{ApJL,515,1999,53}
\bibitem{rf:8}
   S.\ Vaughan, T.\ Boller, A.\ C.\ Fabian, D.\ R.\ Ballantyne, W.\
   N.\ Brandt, \& J.\ Tr\"umper, \JL{MNRAS,337,2002,247}
\bibitem{rf:9}
   T.\ Boller, et al.\ \JL{MNRAS,329,2002,L1}
\bibitem{rf:10}
   K.\ M.\ Leighly, A.\ Zdziarski, T.\  Kawaguchi, \& C.\ Matsumoto,
   2002, Proc.\ ``Workshop on   X-ray Spectroscopy of AGN with Chandra
   and XMM--Newton'', eds.\ Th.\    Boller, S.\ Komossa, S.\ Kahn, H.\
   Kunieda, \& L.\ Gallo (MPE:    Garching) p.\ 259, astro-ph/0205539
\bibitem{rf:11}
   K.\ M.\ Leighly \& J.\ R.\ Moore, submitted to ApJ,
   astro-ph/0402453
\bibitem{rf:12}
   D.\ Proga, J.\ M.\ Stone, \& T.\ R.\ Kallman, \AJ{543,200,686}
\bibitem{rf:13}
   N.\ Murray, J.\ Chiang, S.\ A.\ Grossman, \AJ{451,1995,498}
\bibitem{rf:14}
   K.\ M.\ Leighly, submitted to ApJ,    astro-ph/0402452
\bibitem{rf:15}
   E.\ M.\ Puchnarewicz, K.\ O.\ Mason, \& A.\ Siemiginowska,
   \JL{MNRAS,293,1998,52P}
\bibitem{rf:16}
   D.\ Casebeer, K.\ M.\ Leighly, to be submitted to ApJ
\bibitem{rf:17}
   K.\ M.\ Leighly, J.\ P.\ Halpern, \& E.\ B.\ Jenkins, in Proc.\
   {\it AGN Physics with the Sloan Digital Sky Survey}, eds.\ G.\ T.\
   Richards \& P.\ B.\ Hall (ASP: San Francisco), astro-ph/0402535 



\end{thebibliography}
\end{document}